\input{aipcheck}

\documentclass[
    ,final            
  ]
  {aipproc}

\layoutstyle{8x11single}

\begin{document}

\title{Relativistic interaction of a high intensity photon beam with a plasma: a possible GRB emission mechanism }

\classification{95.85.Nv, 95.85.Pw, 98.70.Rz}
\keywords {X-ray, gamma-ray, gamma-ray sources, gamma-ray bursts}     

\author{G. Barbiellini}{
address={University of Trieste, via Valerio 2, 34100, Trieste, Italy},
altaddress={INFN of Trieste, Padriciano 99, 34012, Trieste, Italy}
}

\author{A. Galli}{
address={INFN of Trieste, Padriciano 99, 34012, Trieste, Italy},
altaddress={IASF of Rome/INAF, Via fosso del cavaliere 100, Roma, 00133, Italy} }

\author{L. Amati}{
address={IASF of Bologna/INAF, Via Gobetti 101, 40129, Bologna, Italy}
}

\author{A. Celotti}{
address={SISSA, via Beirut 2, 34100, Trieste, Italy}
}

\author{R. Landi}{
address={IASF of Bologna/INAF, Via Gobetti 101, 40129, Bologna, Italy}
}

\author{F. Longo}{
address={University of Trieste, via Valerio 2, 34100, Trieste, Italy},
altaddress={INFN of Trieste, Padriciano 99, 34012, Trieste, Italy}
}

\author{N. Omodei}{
address={INFN of Pisa, Edificio C-Polo Fibonacci-Largo B. Pontecorvo 3, Pisa, Italy}
}

\author{M. Tavani}{
address={IASF of Rome/INAF, Via fosso del cavaliere 100, Roma, 00133, Italy} 
}

\begin{abstract}
A long duration photon beam can induce macroscopic coherent effects on a plasma by single photon electron scattering if the probability of the interaction approaches 1 in a volume of unit surface and length equal to the plasma typical wavelength and the induced electron oscillations become relativistic in few plasma cycles. A fraction of the plasma electrons is accelerated through the Wakefield mechanism by the cavities created by the photon-electron interactions and radiates through boosted betraton emission in the same cavities. The resulting emission in this framework is very similar to the typical GRB radiation. Several comparisons with GRB light curves and spectral-energy correlations will be presented.
\end{abstract}

\maketitle
\section{Motivation for an alternative GRB emission mechanism}
\label{introduzione}

An analysis of $\sim$ 400 BATSE GRB light curves revealed the presence of small bumps in the burst count rate to peak count rate ratio at time scale of $10^2$-$10^3$ sec \cite{con02}. \cite{barbie04} ascribed these bumps to the Compton scattering of the prompt gamma-ray photons on the circumburst material surrounding the burst progenitor. \cite{barbie04} assumed a shell of material with uniform density $n$ and thickness $R_0$ located at distance $R_0$ from the central engine, which implies an optical depth to Compton scattering $\tau \sim \sigma n R$. Because of the Compton scattering the prompt gamma-ray photons are attenuated by a factor $e^{-\tau}$, and the resulting synchrotron flux is $L_s \sim (n_p e^{- \tau})/ (\pi \theta_j^2 t_{grb})$,
where $n_p$ is the number of gamma-ray photons, $\theta_j$ is the burst jet aperture, and $t_{grb}$ is the burst duration. The fraction $(1- e^{- \tau})$ of not attenuated photons is Compton scattered by the external material emitting a flux $L_c \sim (n_p (1-e^{- \tau}))/(\pi \theta_j^2 t_{geo})$
where $t_{geo} \sim 2R \theta_j^2/c$ is the time scale of photon diffusion within the shell of material. Then, the ratio between the attenuated synchrotron flux and that resulting by the Compton scattering is $Q=((e^{\tau}-1)c t_{grb})/(2 R)$.
Assuming the shell to be located at $\sim 10^{15}$cm from the central engine, and the jet aperture to be $\theta_j \sim 0.1$, from the observed ratio $Q$ one derives column densities of the order of $10^{24}~cm^{-2}$. These high column densities can be reconciled with that observed for neutral hydrogen only if the surrounding material is 99 $\%$ ionized, and finally imply a surrounding shell of gas of density $n \sim 10^{9}~cm^{-3}$ at $\sim 10^{15}~cm$. 

The compact nature of GRB progenitor and the observed photon spectra require a relativistic motion of the emitting source and photon propagation transparency. This motivated us to search for a mechanism alternative to the fireball able to produce GRB, and that takes advantage of such high densities. We propose that the acceleration mechanism working in GRB is wake-field particle acceleration activated by a very energetic precursor \cite{barbie06}. This idea is supported by recent experimental results, which showed that a laser beam of duration shorter or comparable to the inverse of the plasma frequency, illuminating a gas with density $n \sim 10^{19}~cm^{-3}$, can accelerate electrons up to tens of MeV, and the spectrum of the radiation emitted by these electrons is very similar to that of synchrotron emission \cite{wfa05}. Laboratory measurements also showed that to activate choerent wake-field particle acceleration in a gas with density $n \sim 10^{19}~cm^{-3}$ a power-surface density threshold $\sigma_{Wt}=0.3n~Wcm^{-2}$ has to be overcome. As showed by \cite{barbie06} the power density associated with the GRB precursor is several order of magnitude greater than the power density corresponding to $n \sim 10^{9}~cm^{-3}$.


\section{Theory: Stochastic WFA plasma acceleration}
To realize the Stocastich Wake Field Acceleration (SWFA) regime the interaction probability between the precursor photons and the plasma electrons approaches unity, which  implies \cite{barbie07}:

\begin{equation}
\label{swfacondition}
N_\gamma \frac{\sigma_T \lambda_\Gamma}{c T_{pre} R^2 \theta_j^2} \geq 1
\end{equation}

where $\sigma_T \lambda_\Gamma$ is the interaction volume ($\lambda_\Gamma= (2 \Gamma)^{1/2} \lambda_p$ with $\lambda_p$ the plasma wavelength), $T_{pre}$ is the precursor duration, and $c T_{pre} R^2 \theta_j^2$ is the volume containing the precursor photons. The beam photon number $N_{\gamma}$ is assumed to be $N_\gamma= E_{\gamma,pre}/E_{p,pre}$
with $E_{\gamma,pre}$ and $E_{p,pre}$ the energy and peak energy of the precursor respectively. When the condition \ref{swfacondition} is satisfied electrons acquire a relativistic oscillatory motion in few plasma cycles. These electrons form charged oscillating cavities due to the asymmetry induced in their motion by photon's interactions and the presence within the plasma of positive charged ions. The angular aperture of the radiation emitted by these electrons is related to their stochastic motion. If the real electron path is approximated by a smooth one confined inside the jet cone, the path difference between the electron and the radiated photons is:

\begin{equation}
\label{electrondistance}
R = 24 c T_{pro} / \theta_j^2
\end{equation}

This path difference determines the prompt emission duration $T_{pro}$ of the GRB. The spectrum of the emitted radiation peaks at a typical energy:

\begin{equation}\label{eq:piccospettro}
E_p= \frac{3}{4} h c \gamma^2 \frac{r_0}{\lambda_p^2}
\end{equation}

with $r_0=\lambda_p / \sqrt \gamma$ the impact parameter of the electrons with respect to the positive plasma ions \citep{barbie06}. Because of the stochastic nature of he phenomenon, which is determined by the turbulence of the external medium, the effective peak energy of the emitted photons is obtained correcting eq. \ref{eq:piccospettro} for this turbulence, and it results to be \citep{barbie06}:

\begin{equation}\label{piccoeff}
E_{p,eff} \sim 25 \frac{n}{10^9} \theta_j^{-8/7}~keV
\end{equation}

\section{Prediction: GRB light curves and Spectral-Total Energy correlation}
We now take into account also for the effect of the external medium, which absorption has the effect to attenuate the precursor energy $E_{\gamma,pre}$ by a factor $F(R)$:

\begin{equation}
F(R)=\frac{e^{\sigma n_0 R_0 \big(1-\frac{R}{R_0}\frac{\lambda_0}{\lambda(R)}\big)}}{\big(1+\frac{R}{R_0}\frac{\lambda_0}{\lambda(R)}\big)^2}
\end{equation}

where $R_0$ is the distance from the central engine where the prompt emission is produced. In the prompt emission region the density $n$ of the external medium is roughly constant, thus $\lambda_0$=$\lambda(R)$. Then, using eq. \ref{electrondistance} one finds: 

\begin{equation}
\label{decayprompt}
F(t)=\frac{e^{\sigma n R_0 \big(1-\frac{t}{T_{pro}}\big)}}{R_0^2\big(1+\frac{t}{T_{pro}}\big)^2}
\end{equation}

At times $t > T_{pro}$ the emission intensity decreases exponentially.  When the precursor photons are absorbed by the external medium in such a way that the condition \ref{swfacondition} is not fulfilled, the SWFA regime turns off. At $R >> R_0$ the mean external medium density $n$ starts to decrease, thus the plasma wavelength starts to increases and the SWFA process returns to be efficient producing the afterglow emission. Afterglow emission evolves with time at the same way of the prompt ones (see \cite{barbie07} for details). 
Interestingly, this prediction can provide an explanation for the recent analysis of Swift GRB X-ray light curves performed by \cite{willingale06}, which showed that most of these light curves can be fitted by two components described by the same functional form, an initial exponential decay plus a later power law decay: 

\begin{equation}
\label{decaywill}
F(t) \propto e^{\alpha_c (1-\frac{t}{T_c})},~t<T_c
\end{equation}

\begin{equation}
F(t) \propto \big(\frac{t}{T_c}\big)^{-\alpha_c},~t>T_c
\end{equation}

where $T_c=T_p$ for the first component and $T_c=T_a$ for the second one. $T_p$ and $T_a$ measure the onset of prompt and afterglow emission respectively. 
Comparing our eq. \ref{decayprompt} with eq. \ref{decaywill} for the prompt emission component we find:

\begin{equation}
\label{decayprompt}
\alpha_p= \sigma_T n(R_0) R_0
\end{equation}

With $R_0 \sim 3 \times 10^{15}$cm and $n \sim 10^{9}~cm^{-3}$ we predict $\alpha_p \sim 2$, that is in good agreement with the values of $\alpha_p$ listed in Table 1 of \cite{willingale06}. Analogously for the afterglow:

\begin{equation}
\alpha_a= \frac{1}{2}\sigma_T n_0 R_0 = \frac{1}{2} \alpha_p
\end{equation}

Also this relation between $\alpha_p$ and $\alpha_a$ is in good agreement with the values reported in Table 1 of \cite{willingale06}.
Finally, one can use eq.\ref{decayprompt} with eq. \ref{electrondistance} and the empirical relation implied by the Compton tail of the prompt emission, i.e. $\tau R= \sigma_T n_0 R_0^2$ = const \citep{barbie04}
to express the model parameters $R_0$, $n_0$ and $\theta_j$ in terms of only the measured quantities $\alpha_p$ and $T_{pro}$. This permit us to write GRB peak energy $E_p$ and its isotropic energy $E_{iso}$ in terms of only observed quantities (see \cite{barbie07} for details):

\begin{equation}
\label{epfinale}
E_{p,eff} \propto \big(\frac{T\alpha}{1+z}\big)^{5/7}
\end{equation}

\begin{equation}
\label{eisofinale}
E_{iso,eff} \propto \big(\frac{T\alpha}{1+z}\big)^{27/28}
\end{equation}

We thus predict an  $E_p$-$E_{iso}$ relation (i.e. the Amati relation \cite{amati06}) with index $\sim$0.74, which is higher of about 20 $\%$ from than that reported by \cite{amati06}. However, in the framework of the SWFA model the measured isotropic energy $E_{iso}$  is lower than the predicted one because of the attenuation of the external medium. This attenuation is in good approximation measured by the factor $e^{\alpha(20/E_ {p,keV})}$, with $\alpha = (\alpha_p + 2\alpha_a)/2$. By applying this correction to $E_{iso}$, the measured slope of the $E_p-E_{iso}$ correlation becomes fully consistent with the predicted one.

\section{Conclusions}
In this work we apply the SWFA theory to GRB phenomena, with encouraging results and interesting prospective. In our model GRB are surrounded by a dense material, where the SWFA phenomena takes place. Scaling the plasma density to our values, we obtain a good agreement between our predictions and the typical emission energy of GRBs (i.e $E_p$). We also find that, the development of the typical GRB from the prompt phase to the afterglow phase can be simply explained with a single mechanism whose efficiency depends on the density of the circum-burst medium; moreover we show that exists a threshold above which the SWFA mechanism is effective. Finally, this model naturally predicts the correlation between $E_p$ and the measured isotropic radiated energy $E_{iso}$ (''Amati Relation''), with a slope fully consistent with the measured one \cite{amati06}. This result will be discussed in a future detailed work (in preparation) \cite{barbie07}.

\bibliographystyle{aipproc}   

\bibliography{sample}


\end{document}